\documentclass[twocolumn,showpacs,preprintnumbers,prl]{revtex4}
\usepackage{graphicx,bm,amsmath,amssymb}

\def\gz{\ifmmode{Z\hskip -4.8pt Z}
    \else{\hbox{$Z\hskip -4.8pt Z$}}\fi}

\newcommand{\be}{\begin{equation}}
\newcommand{\ee}{\end{equation}}
\newcommand{\bea}{\begin{eqnarray}}
\newcommand{\eea}{\end{eqnarray}}

\begin{document}

\title{Photoluminiscence of a Quantum Dot Hybridized with a Continuum of Extended States}
\author{L. M. Le\'on Hilario}
\author{A.~A.~Aligia}
\affiliation{Centro At\'{o}mico Bariloche and Instituto Balseiro, Comisi\'{o}n Nacional
de Energ\'{\i}a At\'{o}mica, 8400 Bariloche, Argentina}
\date{\today}

\begin{abstract}
We calculate the intensity of photon emission from a trion in a single
quantum dot, as a function of energy and gate voltage, using the impurity
Anderson model and variational wave functions. Assuming a flat density of
conduction states and constant hybridization energy, the results agree with
the main features observed in recent experiments: non-monotonic dependence
of the energy on gate voltage, non-Lorentzian line shapes, and a line width
that increases near the regions of instability of the single electron final
state to occupations zero or two.
\end{abstract}

\pacs{73.21.La, 78.67.Hc, 75.20.Hr, 72.15.Qm}
\maketitle


Semiconductor quantum dots (QD's) have attracted much attention recently and
have been proposed for numerous applications. It has been established that
they are promising candidates in quantum information processing, with spin
coherence times of several microseconds \cite{pett,grei} and fast optical
initialization and control \cite{ata,grei2,bere}. QD's are also of interest
in the field of spintronics (electronics based on spin) \cite{mac,kork,rei}.
Optical means of manipulation \cite{rei} and detecting \cite{kork} the spin
were proposed. QD's might also be used as semiconductor optical amplifiers 
\cite{gom}. Most of the above cited works involve optical transitions which
create or destroy excitons. In particular, when a positive gate voltage is applied,
configurations with one more electron can be  stabilized, and the optical transition takes place
between a state with the valence band full and only one electron in the conduction 
band, and a state which consists of two conduction electrons and a hole 
in the valence band (a trion)
\cite{ata,grei2,bere,war,hog,smi,dal}.
The gate voltage $V_{g}$ allows to control the spin dynamics and the optical emission 
\cite{war,hog,smi,dal}.

An interesting aspect of the photoluminiscence (PL) decay of the excitons
and trions, is the manifestation of the hybridization of the spin localized
in the QD with a continuum of extended states, and the related Kondo effect 
\cite{smi,dal,helm}. This hybridization can be described by the impurity Anderson
model \cite{var} [see Eq. (\ref{h}) below]. In the limit of small
hybridization and odd number of electron in the QD, 
this model reduces to the Kondo model, in which the localized
spins have an exchange interaction with the spin of the extended states 
\cite{smi}. The Kondo physics is also clearly present in transport properties of
QD's \cite{grobis,rin}.

In recent experiments, the PL that results from the decay of the $X^{1-}$
trion has been measured as a function of $V_{g}$ \cite{dal}. The PL is a
consequence of an optical transition from the trion to a state with one
localized electron in the QD, hybridized with a continuum. As a consequence
of this hybridization, the PL is broad (half width at half maximum $w\sim 1$
meV) near the low $V_{g}$ limit of stability of the trion (for lower $V_{g}$%
, the neutral exciton $X^{0}$ is stable). This limit is slightly below $%
V_{g}^{0}$, where we call $V_{g}^{n}$ the gate voltage for which the
configurations with $n$ and $n+1$ electrons in the QD, and no holes in the valence band,  
are degenerate. The
striking observed behavior can be summarized in three main results: i)
as $\Delta V_{g}=V_{g}-V_{g}^{0}$ increases from negative values, the PL first
blueshifts and then redshifts, ii) the PL line shape is non-Lorentzian, with
a low energy tail, iii) while $w\sim 1$ meV for $V_{g}\sim V_{g}^{0}$ or 
$V_{g}\sim V_{g}^{1}$ (near the limits of stability of one electron in the
QD), the PL line is much narrower for intermediate values of $V_{g}$. This has been
noted before \cite{ata}: while at intermediate gate voltages Atat\"{u}re \textit{et al.}
succeded to prepare a spin state with more than 99.8 \% fidelity,
the coupling of the spin with the reservoir at the edges 
($V_{g}\sim V_{g}^{0}$ or $V_{g}\sim V_{g}^{1}$) spoils this efficiency.  

The line shape has been quantitatively fit only for $V_{g}\sim V_{g}^{0}$ using
a one-particle approach with additional questionable assumptions \cite{dal}. 
In this Letter, we calculate the PL on the whole range
of gate voltages using many-body variational wave functions (VWF's) 
\cite{var,hews}. 
Our results
provide an explanation of the three main experimental results mentioned above
and show that they are a consequence of strong correlations.
We also show that the shift of the PL line is directly related with its
asymmetry by a simple sum rule.

Using Fermi's golden rule, the PL intensity 
at zero temperature
is given by

\begin{equation}
I(\omega )=\frac{2\pi }{\hbar }\sum_{f}|\langle f|H_{LM}|i\rangle|^{2}
\delta (\hbar \omega +E_{f}-E_{i}),  \label{i1}
\end{equation}
where $\omega $ is the PL frequency, $|i\rangle $ is the initial (trion)
state, $|f\rangle $ denotes the possible final states, $E_{j}$ 
is the energy of the state $|j\rangle$, and the relevant part of
light-matter interaction can be written as

\begin{equation}
H_{LM}=(A\sum_{\sigma }d_{\sigma}^{\dagger }p_{\sigma }+\text{H.c.}), \label{hlm} 
\end{equation}
where $A$ is proportional to the vector
potential, $d_{\sigma }^{\dagger }$ creates an electron at the QD (with spin 
$\sigma $), and $p_{\sigma }$ annihilates a valence electron at the QD, with
the same symmetry as that of the corresponding absorbed photon.

The experimental results indicate that the hybridization of the trion state
with the continuum is small. In any case the hybridization 
between holes and extended states favors a singlet
state. Taking the limit of vanishing hybridization  
of the VWF of Varma and Yafet \cite{var} for this 
case [in analogy to Eqs. (\ref{g}), 
(\ref{Eg}) and the discussion below them], one obtains \cite{note3}

\begin{equation}
|i\rangle =\frac{1}{\sqrt{2}}\sum_{\sigma }c_{k_{F}\sigma }^{\dagger
}p_{\sigma }d_{\uparrow }^{\dagger }d_{\downarrow }^{\dagger }|S\rangle ,
\label{ini}
\end{equation}%
where $c_{k \sigma }^{\dagger }$ creates a conduction electron in a
spherical wave with wave vector $k$, $k_{F}$ refers to the Fermi wave vector
and $|S\rangle $ is the many-body state with all valence states of the dot
and all conduction states below the Fermi energy occupied. Replacing Eq. 
(\ref{ini}) in Eq. (\ref{i1}), the intensity becomes

\begin{equation}
I(\omega )=\frac{2\pi }{\hbar }|A|^{2}\rho _{1}(E_{i}-\hbar \omega ),
\label{i2}
\end{equation}
where

\begin{equation}
\rho _{1}(\omega )=\sum_{f}|\langle f|1\rangle |^{2}\delta (\hbar \omega
-E_{f})  \label{rho1}
\end{equation}
is the spectral density of the many-body state 

\begin{equation}
|1\rangle =(d_{\uparrow
}^{\dagger }c_{k_{F}\downarrow }^{\dagger }-d_{\downarrow }^{\dagger
}c_{k_{F}\uparrow }^{\dagger })|S\rangle /\sqrt{2}, \label{1} 
\end{equation}
obtained after the application of $H_{LM}$ [Eq.(\ref{hlm})] on $|i\rangle$ [Eq.(\ref{ini})].

Since in the possible final states $|f\rangle$, the holes are absent and do not play 
any role, our task reduces to
calculate $\rho _{1}$ for the Hamiltonian of the Anderson model, which reads

\begin{eqnarray}
H &=&E_{d}\sum_{\sigma }d_{\sigma }^{\dagger }d_{\sigma }+Ud_{\uparrow
}^{\dagger }d_{\uparrow }d_{\downarrow }^{\dagger }d_{\downarrow
}+\sum_{k\sigma }\epsilon _{k}c_{k\sigma }^{\dagger }c_{k\sigma }  \notag \\
&&+\sum_{k\sigma }(V_{k}d_{\sigma }^{\dagger }c_{k\sigma }+\text{H.c.)},
\label{h}
\end{eqnarray}%
where $E_{d}=E_{d}^{0}-eV_{g}/\lambda $ (with $\lambda \simeq 7$ \cite{dal})
is controlled by the gate voltage $V_{g}$. 

It is easy to see that in the non-interacing case ($U=0$), $\rho _{1}$ is given 
essentially by the spectral density of $d$ electrons, which is a symmetric
Lorentzian for the usual assumptions of constant 
$V_{k}$ and constant density of extended states per
spin $\rho (\omega)$ \cite{ham}. Therefore, to explain the observed
behavior within a one-electron picture, some structure in either 
$V_{k}$ or $\rho (\omega)$ has to be assumed \cite{dal}. Instead, as we show below,
the three main experimental results mentioned above appear naturally 
when the interactions are properly taken into account. 

We solve the problem using many-body VWF's \cite{var,hews}. Similar VWF's
were used successfully to study the interaction between spins of two QD's 
\cite{sim}. We extend previous approaches to the case of finite $U$. To
simplify the discussion, we begin assuming $U\rightarrow \infty $, which is
a good approximation when $E_{d}$ is near the Fermi energy $\epsilon
_{F}=\epsilon _{k_{F}}$, which we take as zero in what follows. 
For $U\rightarrow \infty $, 
the VWF proposed by Varma and Yafet is a good approximation for the
correlated ground state $|g\rangle $ of $H$ \cite{var}

\begin{equation}
|g\rangle =\alpha |F\rangle +\sum_{k\sigma }\frac{\beta _{k}}{\sqrt{2}}%
d_{\sigma }^{\dagger }c_{k\sigma }|F\rangle ,  \label{g}
\end{equation}
where $|F\rangle =c_{k_{F}\uparrow }^{\dagger }c_{k_{F}\downarrow }^{\dagger
}|S\rangle $. Assuming constant $V_{k}$ and 
$\rho(\omega)$, the resulting approximate ground state energy $E_{g}$ comes
from the solution of the equation 

\begin{equation}
E_{d}-\delta +(2\Delta /\pi )\ln \left[(W+\delta )/\delta \right] =0, \label{Eg}
\end{equation}
where $\delta =E_{d}-E_{g}>0$, $\Delta =\pi
\rho V^{2}$ is the resonant level width and $-W$ is the bottom of the band
of extended states \cite{note}. Note that in the limit $V \rightarrow 0$ 
when $E_d < \epsilon_{F}$  Eq.(\ref{g}) gives $|g\rangle = |1\rangle$.
Therefore, the state $|1\rangle $ can be regarded as the limit of the Kondo ground state,
when the Kondo temperature $T_K \sim \text{exp} (\pi E_d/ 2 \Delta) \rightarrow 0$. 

In the thermodynamic limit, $|g\rangle $ is
orthogonal to the state $|1\rangle $. However, it is easy to see that
excited states can be obtained making electron-hole excitations on 
$|g\rangle $, which have energy near $\langle 1|H|1\rangle =E_{d}$ \cite{note}
and hybridize with $|1\rangle $. This hybridization leads to a broadening of 
$\rho _{1}(\omega )$, in a way analogous to a resonant level model, in which
a localized state is mixed with a continuum \cite{ham}. Specifically we
consider the states

\begin{equation}
|k^{\prime }\rangle  = \alpha ^{k^{\prime }}|e_{k^{\prime }}\rangle
+\sum_{k}\beta _{k}^{k^{\prime }}|y_{kk^{\prime }}\rangle , \label{kpri}
\end{equation}
where

\begin{eqnarray}
|e_{k^{\prime }}\rangle  &=&\frac{1}{\sqrt{2}}(c_{k^{\prime }\uparrow
}^{\dagger }c_{k_{F}\downarrow }^{\dagger }-c_{k^{\prime }\downarrow
}^{\dagger }c_{k_{F}\uparrow }^{\dagger })|S\rangle ,  \notag \\
|y_{kk^{\prime }}\rangle  &=&\frac{1}{\sqrt{2}}\sum_{\sigma }d_{\sigma
}^{\dagger }c_{k\sigma }|e_{k^{\prime }}\rangle .  \label{ey}
\end{eqnarray}
Here the subscript $k$ ($k^{\prime }$) denotes extended states below (above)
the Fermi energy. It is easy to see that optimization of the coefficients
for each $k^{\prime }$ (minimizing the energy), leads to $\alpha ^{k^{\prime
}}=\alpha $, $\beta _{k}^{k^{\prime }}=\beta _{k}$, independently of 
$k^{\prime }$. In addition, $|\langle 1|H|k^{\prime }\rangle |^{2}=|\alpha
V_{k^{\prime }}|^{2}=(1-n_{d})|V_{k^{\prime }}|^{2}$, 
where $n_{d}=$ $\langle g|\sum_{\sigma }d_{\sigma }^{\dagger
}d_{\sigma }|g\rangle $ is the electronic occupation at the QD. From the
effective resonant level model, one obtains a broadening of the PL line
given approximately by $w=$ $(1-n_{d})\Delta $. This is a simple, elegant
result which agrees qualitatively with experiment near the intermediate
valence regime (IVR) $|E_{d}-\epsilon _{F}|\sim \Delta $, and predicts an
exponentially small width in the Kondo regime $\epsilon
_{F}-E_{d},E_{d}+U-\epsilon _{F}\gg \Delta $. However, since double
occupancy is neglected, this result is clearly wrong in the other IVR $%
|E_{d}+U-\epsilon _{F}|\sim \Delta $, while in the Kondo regime, virtual
fluctuations through double occupied states were neglected.

To improve this calculation, we first replace the states  $|y_{kk^{\prime
}}\rangle $ in Eq. (\ref{kpri}) by others $|\tilde{y}_{kk^{\prime }}\rangle $%
, which take into account double occupancy

\begin{equation}
|\tilde{y}_{kk^{\prime }}\rangle =\gamma |y_{kk^{\prime }}\rangle
+\sum_{q\sigma }\frac{\beta _{q}}{\sqrt{2}}d_{\sigma }^{\dagger }c_{q\sigma
}|y_{kk^{\prime }}\rangle , \label{it}
\end{equation}
where the coefficients are determined minimizing $\langle \tilde{y}%
_{kk^{\prime }}|H|\tilde{y}_{kk^{\prime }}\rangle $, for fixed $k$ and $%
k^{\prime }$. Second, in a similar way to the resulting states $|k^{\prime
}\rangle $, we construct linear combinations $|k\rangle $ appropriate for
the other IVR  $|E_{d}+U-\epsilon _{F}|\sim \Delta $. This is done easily
using a symmetry property of $H$: 
$d_{\uparrow }^{\dagger }\rightarrow d_{\downarrow }$, $c_{k\uparrow }^{\dagger
}\rightarrow -c_{k^{\prime }\downarrow }$, 
$d_{\downarrow }^{\dagger }\rightarrow -d_{\uparrow }$, $c_{k\downarrow
}^{\dagger }\rightarrow c_{k^{\prime }\uparrow }$, with $\epsilon
_{k^{\prime }}=-\epsilon _{k}$ (for $\epsilon _{F}=0$), which changes -$E_{d}$ into $E_{d}+U$ and
therefore interchanges IVR's. 
For example using this transformation $|e_{k^{\prime }}\rangle
\rightarrow d_{\uparrow }^{\dagger }d_{\downarrow }^{\dagger }\sum_{\sigma
}c_{k_{F}\sigma }^{\dagger }c_{k\sigma }|S\rangle /\sqrt{2}$. Third, since
the states $|k^{\prime }\rangle $ and $|k\rangle $ are not orthogonal, we
discretize them and use a symmetric L\"{o}wdin procedure to orthogonalize them
\cite{low}. Finally, including the state $|1\rangle $, we numerically
diagonalize $H$ and calculate the spectral density using Eq. (\ref{rho1}).

\begin{figure}[tbp]
\includegraphics[width=7.5cm]{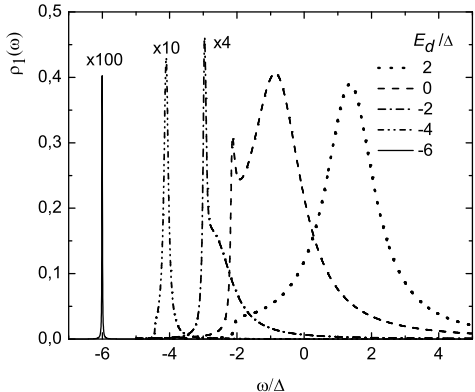}
\caption{density of the state $|1\rangle$ as a function of frequency for
several values of $E_d$. For the lower values of $E_d$ the amplitude was reduced
by the factors indicated in the figure.}
\label{dens}
\end{figure}

The parameters were chosen as follows: $\Delta $ is our unit of energy. It
should de of the order of 1 meV to give the observed PL width for $E_{d}\sim
\epsilon _{F}$. Guided from experiment \cite{dal}, we took $U=24\Delta $.
Finally we took a broad flat band symmetrically placed around $\epsilon _{F}$
of width $2W=100\Delta $. The results depend only logarithmically on $W$ 
[see for instance Eq. (\ref{Eg})]
and the band edges are very far form the region of interest. Therefore there
are no particular features of the band or hybridization that can affect the
results. To discretize the states $|k^{\prime }\rangle $ and $|k\rangle $,
the band was divided in at least 2000 points with decreasing energy
separation towards the Fermi energy \cite{note2}.

In Fig. \ref{dens} we represent $\rho _{1}(\omega )$ 
as a function of frequency $\omega$ for various values of $E_{d}$. This density
with the frequency axis inverted and shifted by the energy of the initial
state $E_{i}-E_{S}$ \cite{note} corresponds to the PL spectrum [see Eq. (\ref%
{i2})]. Due to our choice of a symmetric band, and the above mentioned
symmetry property of $H$, the densities for  $E_{d}$ and $-E_{d}-U$ are
identical. Therefore, one can restrict the study to  $E_{d}>-U/2=-12\Delta $%
. Note that in most curves an asymmetry with a high energy tail
(corresponding to a low energy tail in the PL) is evident, in agreement with
experiment. We return to this point below. For $E_{d}=0$ or slightly
negative, there are two relative maxima in  $\rho _{1}(\omega )$. 
This structure is already present using the simplest VWF's given by 
Eqs. (\ref{kpri}) and (\ref{ey}).
The peak
at smallest energies corresponds to excitations near the ground state. 
At low energies, one expects that $\rho _{1}(\omega )$ has a power-law
behavior characteristic of x-ray edge singularities \cite{helm,noz} which is not
captured by our variational approach. 
This structure might be washed 
by a temperature or other broadening effects not
included in our calculations. This is likely the reason why a structure with
two peaks has not been so far reported.
Note that while the density near the ground-state energy
can be described more accurately with the numerical
renormalization group \cite{helm,bulla}, this technique does not resolve
accurately the structure at higher energies \cite{bulla,vau}.  

\begin{figure}[tbp]
\includegraphics[width=7.0cm]{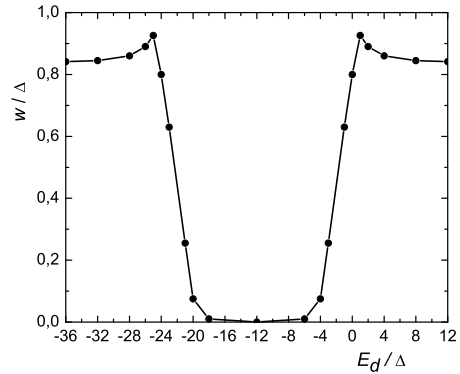}
\caption{Half width at half maximum of the density of the state $|1\rangle$ 
as a function of $E_d$.}
\label{w}
\end{figure}

One striking feature observed in Fig. \ref{dens} is the dramatic narrowing
of the peak as the system leaves the IVR and enters the Kondo regime. in
Fig. \ref{w} we represent the half width at half maximum $w$ as a function
of the dot level, which is related to the gate voltage by $E_{d}=-e\Delta
V_{g}/7$. The slight increase of $w$ for $\Delta V_{g}=0$ (and also $%
E_{d}=-24\Delta $) might be related to the two-peak structure of $\rho
_{1}(\omega )$ near these values of $E_{d}$ and we believe it is not too
significative. Instead, for the symmetric case $E_{d}=-12\Delta $, we
obtain $w=5\times 10^{-4}\Delta $, corresponding to a decrease of the line width 
in three orders of magnitude. This is consistent with different experiments \cite{ata,hog,dal}. 

\begin{figure}[tbp]
\includegraphics[width=6.5cm]{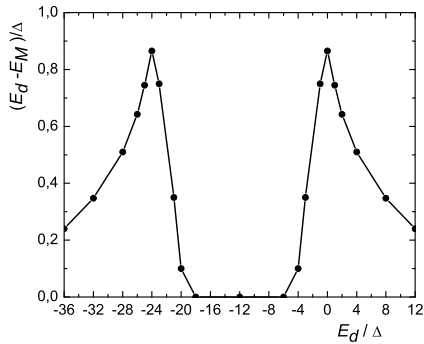}
\caption{Shift of the maximum of the density of the state $|1\rangle$ 
as a function of $E_d$.}
\label{shift}
\end{figure}

In Fig. \ref{shift} we plot $E_{M}$, the energy $\hbar \omega $ for which $%
\rho _{1}(\omega )$ reaches its maximum value as a function of $%
E_{d}=-e\Delta V_{g}/7$. The shift $E_{M}-E_{d}$ is always negative (the
peak is displaced towards the ground state energy). This means that the
energy of the final state decreases and therefore, the position of the PL
peak is always blueshifted with respect to the situation without
hybridization. However, the non-monotonic behavior observed in Fig. \ref%
{shift} implies that as $\Delta V_{g}$ increases starting from negative
values, the PL peak first blueshifts, but after $\Delta V_{g}=0$, the peak
redshifts, in agreement with experiment \cite{dal}. The fact that the maximum
shift is obtained near the IVR's might be expected from the fact the maximum 
energy gain in the ground state is obtained at those regimes. 

It is easy to see from Eq. (\ref{rho1}) that $\int d(\hbar \omega )\rho
_{1}(\omega )=\langle 1|H|1\rangle =E_{d}$ \cite{note}. In words, the
center of gravity of the density states remains at $E_{d}$ 
when the QD electron hybridizes, while its maximum
displaces at a lower energy $E_{M}$. Therefore, to satisfy the sum rule, 
$\rho _{1}(\omega )$ should be asymmetric with a high energy tail, and the
degree of asymmetry is directly proportional to the magnitude of the shift 
$|E_{M}-E_{d}|.$

In summary, using an impurity Anderson model to describe the final states of
a photoluminiscence experiment in a quantum dot, with the simplest possible
assumptions for the density of extended states and their hybridization with the dot electrons, we
obtain a blueshift and correlated asymmetry of the line shape, and a line
width, which explains the main features of recent experiments in the whole range of 
applied gate voltage.

We thank R. J. Warburton and D. Garc\'{\i}a for useful discussions. We are supported by
CONICET. This work was done in the framework of projects PIP 5254 of
CONICET, and PICT 2006/483  of the ANPCyT.


\begin{thebibliography}{99}
\bibitem{note3} For simplicity we neglect the spin-orbit interaction of the
hole states. Including this interaction and crystal-field effects \cite{ata}, 
the initial state
is more properly described by 
$|i\rangle =c_{k_{F}\uparrow }^{\dagger
}p_{1 \uparrow }d_{\uparrow }^{\dagger }d_{\downarrow }^{\dagger }|S\rangle$, 
where $p_{1 \uparrow }$ annihilates a valence electron with maximum angular momentum 
and spin projections in the growth direction (or opposite to it).
This however leaves the results unchanged within our approach.
\bibitem{note} The energy of the state $|S\rangle $ ($E_{S}$) enters as an
additive constant in the energy of all many-body states considered here and
is dropped for simplicity.
\bibitem{note2} Succesive intervals between $\epsilon _{k}$ are multiplied
by a factor $1/\Lambda $, with $\Lambda >1$, as $\epsilon _{k}$ approaches 
$\epsilon _{F}$, keeping $\rho V_{k}^{2}$ constant \cite{bulla}. $\Lambda $
is optimized depending on $E_{d}$.
\end{thebibliography}
\end{document}